\documentclass[prl,aps,10pt,superscriptaddress,twocolumn,floatfix]{revtex4-2}

\usepackage[nice]{nicefrac}
\usepackage{graphicx,color,comment}
\usepackage{amsmath,amssymb,bm,bbm}
\usepackage{amsmath,upgreek}
\usepackage{braket}
\usepackage{bbold}
\usepackage{float}
\usepackage{physics}

\DeclareMathOperator\arctanh{arctanh}

\newcommand{\bb}{\kappa}
\newcommand{\ie}{{\it i.e.},\ }

\usepackage[plainpages=false,pdfpagelabels,colorlinks=true,linkcolor=red,urlcolor=blue,citecolor=blue,pdftitle={},pdfauthor={},pdfdisplaydoctitle=true,pdfduplex=DuplexFlipLongEdge]{hyperref}

\definecolor{darkred}{rgb}{0.90,0.2,0.2}
\definecolor{darkgreen}{rgb}{0,0.60,.2}
\definecolor{darkblue}{rgb}{0.1,0.3,1}
\definecolor{grey}{cmyk}{0,0,0,0.25}
\definecolor{orange}{cmyk}{0,0.6,0.8,0}

\usepackage{appendix}

\usepackage[T1]{fontenc}
\usepackage{orcidlink}
\begin{document}

\title{One-Body~Purity,~Non-Gaussianity,~and~Entanglement~in~Interacting~Integrable~Models}

\author{R. Świętek\orcidlink{0009-0004-5353-9998}}
\affiliation{Institut f\"ur Theoretische Physik, Georg-August-Universität G\"ottingen, D-37077 G\"ottingen, Germany\looseness=-1}
\author{M. Kliczkowski\orcidlink{0000-0002-7987-9913}}
\affiliation{Center for Advanced Systems Understanding, Helmholtz-Zentrum Dresden-Rossendorf, Germany\looseness=-3}
\affiliation{Institute of Theoretical Physics, Faculty of Fundamental Problems of Technology, Wrocław University of Science and Technology, 50-370 Wrocław, Poland\looseness=-3}
\author{L. Vidmar\orcidlink{0000-0002-6641-6653}}
\affiliation{Department of Theoretical Physics, J. Stefan Institute, SI-1000 Ljubljana, Slovenia\looseness=-1}
\affiliation{Department of Physics, Faculty of Mathematics and Physics, University of Ljubljana, SI-1000 Ljubljana, Slovenia\looseness=-1}
\author{M. Rigol\orcidlink{0000-0002-5806-5873}}
\affiliation{Department of Physics, The Pennsylvania State University, University Park, Pennsylvania 16802, USA}

\begin{abstract}
When describing entanglement in typical midspectrum eigenstates of many-body lattice Hamiltonians, two paradigms have emerged that capture the behavior observed in integrable and nonintegrable systems, Haar-random fermionic Gaussian states and Haar-random pure states, respectively. Remarkably, the former capture the behavior of interacting integrable systems, whose eigenstates are non-Gaussian. We argue that the paradigm that captures both the entanglement properties and the lack of Gaussianity in integrable systems is that of random superpositions of polynomially many Gaussian states. In contrast, eigenstates of nonintegrable systems are consistent with being described by random superpositions of exponentially many Gaussian states. We gain this understanding by comparing analytical and numerical results for the one-body purity, the non-Gaussianity, and the entanglement entropy of the random superpositions and the Hamiltonian eigenstates.
\end{abstract}

\maketitle

\textit{Introduction.} Entanglement has emerged as a powerful probe for quantum many-body systems, linking quantum information concepts to condensed-matter phenomena such as exotic states of matter~\cite{Kitaev2005, amico_vlatko_2008, Calabrese_2009, eisert_cramer_10, laflorencie_quantumentanglementcondensed_2016}, transport~\cite{kim_huse_14}, quantum criticality~\cite{osterloh_2002, osborne_2002, vidmar_hackl_18, hackl_vidmar_19} and thermalization~\cite{popescu_short_06, winter_linden_2009, Calabrese_2005, calabrese07}. Entanglement has also emerged as a sensitive probe of integrability and nonintegrability~\cite{leblond_mallayya_19, bianchi_hackl_22}. Typical midspectrum eigenstates of Hamiltonians with finite-dimensional local Hilbert spaces exhibit a leading volume-law contribution to the bipartite entanglement entropy that is maximal for nonintegrable systems (see Refs.~\cite{bianchi_hackl_22, kliczkowski_swietek_23} and references therein) and submaximal for interacting integrable systems~\cite{leblond_mallayya_19, patil_rigol_23, swietek_kliczkowski_24, yauk_patil_24}. In addition, the normalized standard deviation of the entanglement entropy of energy eigenstates decays exponentially versus polynomially with the volume in nonintegrable versus integrable systems~\cite{swietek_kliczkowski_24}.

Analytical insights into those distinctive entanglement-entropy behaviors have been gained using Haar-random states in the absence~\cite{Page93} and presence of continuous symmetries such as $U(1)$~\cite{vidmar_rigol_17, garrison_grover_18, bianchi_dona_19, yauk_patil_24} and $SU(2)$~\cite{patil_rigol_23, bianchi_dona_24, chakraborty2025} for nonintegrable systems, and Haar-random fermionic Gaussian states~\cite{lydzba_rigol_20,*lydzba_rigol_21, bianchi_hackl_21} for integrable qubit-based systems (our focus here). Haar-random fermionic Gaussian states exhibit a typical volume-law coefficient that depends on the ratio between the volumes of the subsystem and the entire system~\cite{lydzba_rigol_20, *lydzba_rigol_21, bianchi_hackl_21, bianchi_hackl_22}. Remarkably, the numerical results for the typical entanglement entropies of eigenstates of spin-$\tfrac12$ interacting integrable models (which are non-Gaussian) are very close to those predicted analytically for Haar-random fermionic Gaussian states~\cite{leblond_mallayya_19}. Similarly, the normalized standard deviation of the entanglement entropy decays polynomially in both cases~\cite{swietek_kliczkowski_24}. Our goal is to understand why this is so.

Our study also illuminates why the eigenstate fluctuations of few-body observables display a parallel behavior. While  in nonintegrable systems the eigenstate thermalization hypothesis (ETH), whose origins lie in random matrix theory (RMT), predicts exponentially small (in the volume) fluctuations of diagonal matrix elements~\cite{Deutsch91, srednicki_94, rigol_dunjko_08, dalessio_kafri_16}, quadratic~\cite{biroli_kollath_10, lydzba_swietek_24} and interacting integrable~\cite{Ikeda2013, steinigeweg_herbrych_13, beugeling_moessner_14, alba_15, leblond_mallayya_19, Mierzejewski_2020, leblond_rigol_20, zhang_vidmar_22, essler_klerk23} models exhibit only a polynomial suppression.

%%%%%%%%%%%%%%%%%%%%%%
\begin{figure}[!t]
    \includegraphics[width=0.95\columnwidth]{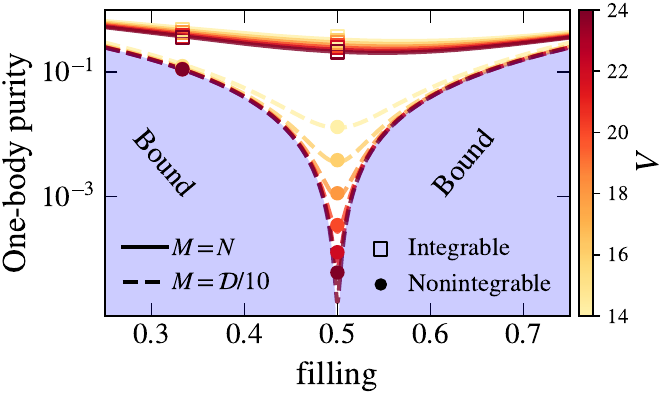}
    \vspace{-0.2cm}
    \caption{One-body purities vs the filling $n\!=\!N/V$ in lattices with $V$ sites (see the color scale) and $N$ fermions. The region below the lower bound from Eq.~\eqref{eq:purity} is shaded in purple. The solid (dashed) lines show the purities obtained for random superpositions of $M\!=\!N$ ($M\!\approx\!\mathcal{D}/10$, where $\mathcal{D}$ is the Hilbert space dimension) fermionic Gaussian states. Empty squares (filled circles) show results for the integrable (nonintegrable) Hamiltonian~\eqref{eq:model} with $\Lambda\!=\!0$ ($\Lambda\!=\!2$).}\label{figM0}
\end{figure}
%%%%%%%%%%%%%%%%%%%%%%

In this Letter, we argue that properties such as the one-body purity, the non-Gaussianity, and the entanglement entropy of typical eigenstates of interacting integrable spinless-fermion lattice systems are well described by random superpositions of $M\!=\!O(N)$ fermionic Gaussian states (where $N$ is the number of fermions), leading to a polynomial Gaussian rank (or Slater rank), which underlies the results reviewed so far. In contrast, to write eigenstates of nonintegrable systems in terms of fermionic Gaussian states, one needs an exponential number of the latter $M\!=\!O(\mathcal{D})$, where $\mathcal{D}$ is the Hilbert space dimension. We support this understanding by comparing analytical and numerical results for the one-body purity, the non-Gaussianity, and the entanglement entropy of random superpositions of fermionic Gaussian states and of eigenstates of spinless fermions at and away from integrability. Our main findings are summarized in Fig.~\ref{figM0}.

\textit{Gaussian~states.}~Bases of fermionic Gaussian states $\{\ket{\uppsi_{\alpha}^G}\}$ naturally appear as the eigenstates of quadratic Hamiltonians $\hat{H}^{}_{\rm qd}$, $\hat{H}^{}_{\rm qd}\ket{\uppsi_{\alpha}^G}\!=\!E^{}_\alpha\ket{\uppsi_{\alpha}^G}$. We focus on $U(1)$ symmetric models so that, using Wick's theorem, all correlation functions for $\ket{\uppsi_{\alpha}^G}$ can be computed in terms of the one-body density matrix (OBDM) $C^{\alpha}_{\ell\ell'}\!=\!\mel{\uppsi_{\alpha}^G}{\hat{c}^\dagger_\ell\hat{c}^{}_{\ell'}}{\uppsi_{\alpha}^G}$~\cite{hackl_bianchi_21}. For analytic calculations, it is convenient to work with the linear complex structure~\cite{hackl_bianchi_21}:
\begin{equation}\label{eq:complex_structure:def}
    (J_{\uppsi_{\alpha}^G})^{}_{\ell\ell'}=\mel{\uppsi_{\alpha}^G}{[\hat{c}^\dagger_\ell,\hat{c}^{}_{\ell'}]}{\uppsi_{\alpha}^G}=2C^{\alpha}_{\ell\ell'}-\delta^{}_{\ell\ell'}\;.
\end{equation}
In terms of the quasiparticle operators $\hat{f}^{}_q\!=\!\sum_\ell v^{}_{\ell q}\hat{c}^{}_\ell$, in which $\hat{H}^{}_{\rm qd}$ is diagonal, the matrix elements of ${\bf J}_{\uppsi_{\alpha}^G}$ can be written as $(J_{\uppsi_{\alpha}^G})^{}_{\ell\ell'}\!=\!\sum_{q=1}^V \text{N}^{\alpha}_q v_{\ell q}^* v^{}_{\ell'\! q}$, where $\text{N}^{\alpha}_q\!=\!2\text{n}^{\alpha}_q-1$ are the rescaled occupations $\text{n}^{\alpha}_q$ of orbital $q$ in the state $\ket{\uppsi_{\alpha}^G}$. The matrices ${\bf J}_{\uppsi_{\alpha}^G}$ have been extensively used in entanglement entropy calculations~\cite{vidmar_hackl_17, vidmar_hackl_18, hackl_vidmar_19, bianchi_hackl_22}.

\textit{Non-Gaussian~states.}~We study two sets of states.
First, we consider random superpositions of fermionic Gaussian states
\begin{equation}\label{eq:superposition}
    \ket{\Psi}=\sum_{\alpha=1}^M a^{}_\alpha\ket{\uppsi_{\alpha}^G},
\end{equation}
which are obtained by drawing independent real (GOE case) or complex (GUE case) Gaussian random variables $\tilde{a}^{}_\alpha$ and then normalizing them to obtain the coefficients $a^{}_\alpha$ satisfying $\sum_\alpha\abs{a^{}_\alpha}^2\!=\!1$. The squared magnitudes $\qty(|a^{}_1|^2,...,|a^{}_M|^2)$ follow a Dirichlet distribution with parameters $(\bb,...,\bb)$, where $\bb\!=\!\frac{1}{2}$ for GOE and $\bb\!=\!1$ for GUE (see Ref.~\cite{SM}), which allows us to perform the calculations analytically. We also consider eigenstates $\ket{\psi^{}_m}$ of many-body interacting spinless-fermions Hamiltonians $\hat H$, $\hat H \ket{\psi^{}_m} \!=\! E^{}_m \ket{\psi^{}_m}$.

We then compute their associated complex structures
\begin{equation}\label{eq:obdm:mixed}
    (J_{\Psi})^{}_{\ell\ell'}=\mel{\Psi}{[\hat{c}_\ell^\dagger,\hat{c}^{}_{\ell'}]}{\Psi},\ \ \ (J_{\psi^{}_m})^{}_{\ell\ell'}=\mel{\psi^{}_m}{[\hat{c}_\ell^\dagger,\hat{c}^{}_{\ell'}]}{\psi^{}_m}\!,
\end{equation}
to evaluate indicators that probe how they differ from fermionic Gaussian states and from each other.

\textit{One-body purity.} As a first indicator, we consider:
\begin{equation}\label{eq:purity:def}
    \ev{\mathcal{P}}^{}_\bb = \frac{1}{V}\overline{\Tr[({\bf J}_{\Psi})^2]}, \ \ \ \bar{\mathcal{P}} = \frac{1}{V}\overline{\Tr[({\bf J}_{\psi^{}_m})^2]},
\end{equation}
which we call the {\bf one-body purities}. The former is calculated analytically, whereas the latter is computed numerically. For fermionic Gaussian states, the one-body purity is $\mathcal{P}\!=\!1$; otherwise, $\mathcal{P}\!<\!1$, and for maximally mixed states $\mathcal{P}\!=\!(2n-1)^2$. For states with fixed number of particles, the one-body purity $\mathcal{P}$ is related to the $k\!=\!1$ fermionic antiflatness recently studied in Ref.~\cite{sierant_turkeshi_26}, see Ref.~\cite{SM}.

Using the rotational invariance of the distribution of coefficients $a^{}_\alpha$, we find an exact expression for the average purity (see End Matter for the derivation):
\begin{equation}\label{eq:purity}
 \ev{\mathcal{P}}^{}_\bb = (2n-1)^2 + 4n(1-n)\!\left[\frac{\bb+1}{M\bb+1} \! \left(1-\frac{V}{\mathcal{D}}\right) \! + \frac{V}{\mathcal{D}}\right],
%    \ev{\mathcal{P}}^{}_\bb = & (2n-1)^2 + 4n(1-n)\frac{\bb+1}{M\bb+1}\nonumber \\ &+\frac{4Vn(1-n)}{\mathcal{D}}\qty[1-\frac{\bb+1}{M\bb+1}].
\end{equation}
which does not depend on the basis $\{\ket{\uppsi_{\alpha}^G}\}$. It depends on the filling $n\!=\!N/V$, the number of lattice sites $V$, the Hilbert-space dimension $\mathcal{D}$, and the ensemble used.

\begin{figure}[!t]
    \includegraphics[width=\columnwidth]{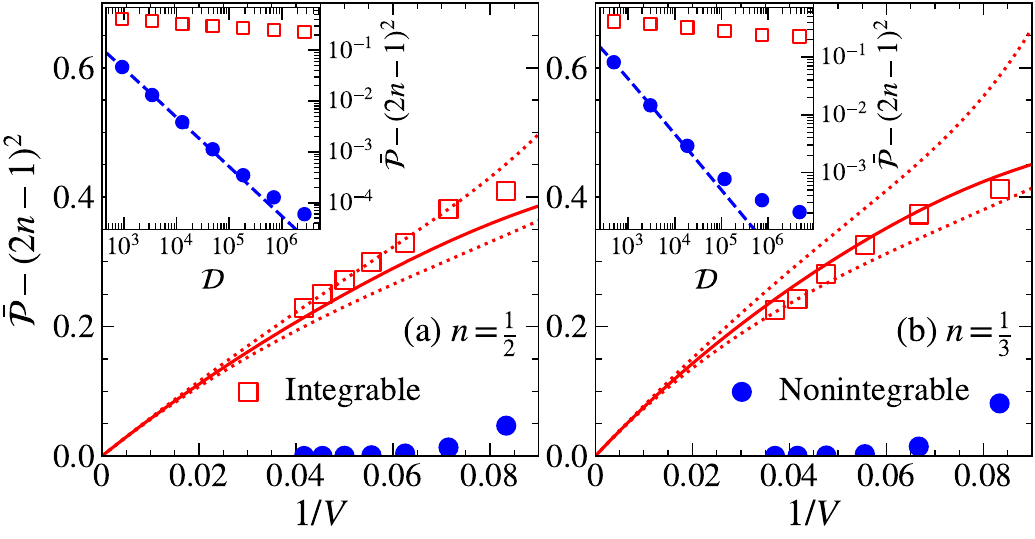}
    \vspace{-0.6cm}
    \caption{Average one-body purity $\bar{\mathcal{P}}-(2n-1)^2$ vs $1/V$ in eigenstates of integrable (nonintegrable) interacting fermions with $\Lambda\!=\!0$ ($\Lambda\!=\!2$) for (a) $n\!=\!\frac{1}{2}$ and (b) $n\!=\!\frac{1}{3}$. The solid (dotted) lines show the predictions of Eq.~\eqref{eq:purity} for $\bb\!=\!1/2$ and $M\!=\!N$ ($M\!=\!N\pm1$). The insets show $\bar{\mathcal{P}}$ vs $\mathcal{D}$. The dashed lines are the predictions of Eq.~\eqref{eq:purity} for $\bb\!=\!1/2$ and $M\!\approx\!0.1\mathcal{D}$.}\label{figM2}
\end{figure}

Equation~\eqref{eq:purity} shows that $\ev{\mathcal{P}}^{}_\bb \!\geq\! (2n-1)^2$, see Fig.~\ref{figM0}. For large system sizes, for which $V\!\ll\! \mathcal{D}$, and large $M\!\gg\!1$ the approach to the lower bound $(2n-1)^2$ is $\propto\!1/M$. In Fig.~\ref{figM2} we show how the bound is approached for $n\!=\!\tfrac12$ [Fig.~\ref{figM2}(a)] and $n\!=\!\tfrac13$ [Fig.~\ref{figM2}(b)] for $M\!=\!N$ and $M\!=\!N\pm1$ (solid and dotted lines, respectively) and $M\!\approx\!0.1\mathcal{D}$ (dashed lines in the insets).

%%%%%%%%%%%%%%%%%%%%%%
%%%%%%%%%%%%%%%%%%%%%%

\textit{Non-Gaussianity.}
As a second indicator, we consider:
\begin{equation}\label{eq:non_gauss:def}
    \mathcal{N} = -\frac{1}{V}\sum_{i=1}^V\qty[g\qty(\frac{1+\nu^{}_i}{2}) + g\qty(\frac{1-\nu^{}_i}{2})]\;,
\end{equation}
which we call the {\bf non-Gaussianity}, where $g(x) \!=\! x\ln x$, and $\nu^{}_i$ are the eigenvalues of ${\bf J}_{\Psi}$ and ${\bf J}_{\psi^{}_m}$ in Eq.~\eqref{eq:obdm:mixed}, and then we compute the corresponding averages $\ev{\mathcal{N}}^{}_\bb$ and $\bar{\mathcal{N}}$, respectively. For a fermionic Gaussian state $\mathcal{N}\!=\!0$ because $\nu^{}_i\!=\!\pm1$, while $\mathcal{N}\!>\!0$ for a non-Gaussian state. Note that the $\tfrac{1}{V}$ normalization in Eq.~\eqref{eq:non_gauss:def} means that $\mathcal{N}$ is the coefficient of the volume scaling of the sum.

%%%%%%%%%%%%%%%%%%%%%%
\begin{figure}[!t]
    \includegraphics[width=\columnwidth]{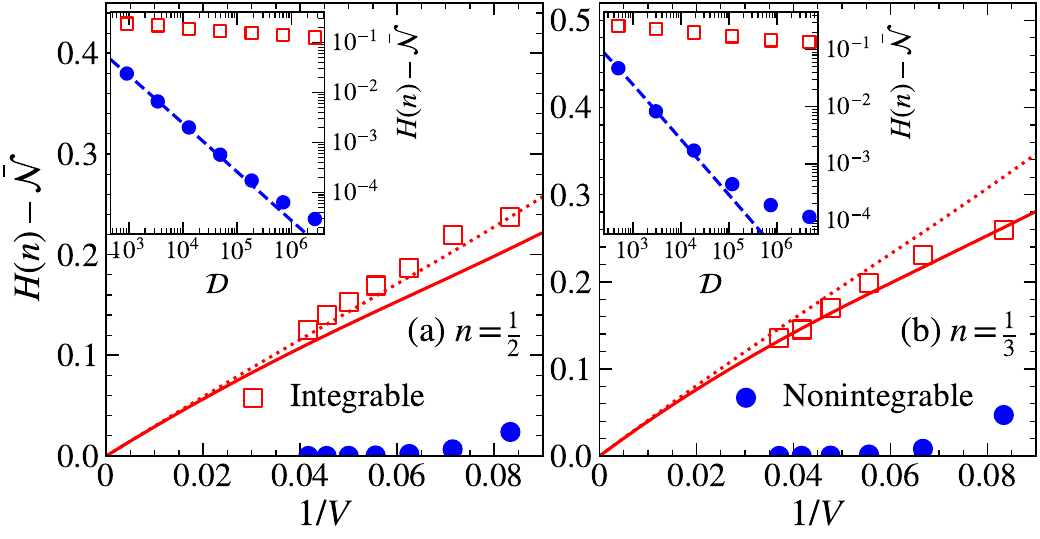}
    \vspace{-0.5cm}
    \caption{Average non-Gaussianity $H(n)-\bar{\mathcal{N}}$ vs $1/V$ in eigenstates of integrable (nonintegrable) interacting fermions with $\Lambda\!=\!0$ ($\Lambda\!=\!2$) for (a) $n\!=\!\frac{1}{2}$ and (b) $n\!=\!\frac{1}{3}$. The solid (dotted) lines show the predictions of Eq.~\eqref{eq:non-gauss} for $\bb\!=\!1/2$ and $M\!=\!N$ ($M\!=\!N-1$). The insets show $\bar{\mathcal{N}}$ vs $\mathcal{D}$. The dashed lines are the predictions of Eq.~\eqref{eq:non-gauss} for $\bb\!=\!1/2$ and $M\!\approx\!0.1\mathcal{D}$.}\label{figM3}
\end{figure}
%%%%%%%%%%%%%%%%%%%%%%

Using a factorization assumption (see End Matter), which becomes exact in the limit of large M (as we verify numerically, see Ref.~\cite{SM}), we obtain a closed-form expression for $\ev{\mathcal{N}}^{}_\bb$ of the form:
\begin{equation}\label{eq:non-gauss}
    \ev{\mathcal{N}}^{}_\bb = \ln{2} - \sqrt{\ev{\mathcal{P}}^{}_\bb}\arctanh{\sqrt{\ev{\mathcal{P}}^{}_\bb}}-\frac{1}{2}\ln(1-\ev{\mathcal{P}}^{}_\bb),
\end{equation}
where the purity $\ev{\mathcal{P}}^{}_\bb$ is given by Eq.~\eqref{eq:purity}. Note that, for $M\!=\!1$, Eqs.~\eqref{eq:purity} and~\eqref{eq:non-gauss} reproduce the result for a fermionic Gaussian state, $\ev{\mathcal{P}}^{}_\bb\!=\!1$ and $\ev{\mathcal{N}}^{}_\bb\!=\!0$. Furthermore, the lower bound for the purity in Eq.~\eqref{eq:purity} implies an upper bound for the non-Gaussianity,
\begin{equation}\label{eq:non-gauss:leading}
    \ev{\mathcal{N}}^{}_\bb\leq H(n)\equiv-n\ln(n)-(1-n)\ln(1-n)\;.
\end{equation}
Remarkably, $\ev{\mathcal{N}}^{}_\bb$ is bounded by the coefficient $H(n)$ of the volume in the entanglement entropy of random pure states with $U(1)$ symmetry~\cite{vidmar_rigol_17, garrison_grover_18, bianchi_hackl_22}, see End Matter. In Fig.~\ref{figM3} we show how $ H(n)$ is approached for $n\!=\!\tfrac12$ [Fig.~\ref{figM3}(a)] and $n\!=\!\tfrac13$ [Fig.~\ref{figM3}(b)] for $M\!=\!N$ and $M\!=\!N-1$ (solid and dotted lines, respectively) and $M\!\approx\!0.1\mathcal{D}$ (dashed lines in the insets).

\textit{Entanglement entropy.} Given a pure state $\ket\Phi$, the bipartite entanglement entropy of a subsystem $A$ comprising $V^{}_A$ sites (which we take to be contiguous sites in our local models) is computed from the reduced density matrix $\hat{\rho}^{}_A\!=\!\Tr^{}_B\dyad{\Phi}$, obtained by tracing out its complement $B$ with $V^{}_B\!=\!V-V^{}_A$ sites. The von Neumann entanglement entropy is then defined as $S^{}_A\!=\!-\Tr(\hat{\rho}^{}_A\ln{\hat{\rho}^{}_A})$. In contrast to $\mathcal{P}$ and $\mathcal{N}$, the analytic calculation of $S^{}_A$ for fermionic Gaussian states depends on the specific states considered and is very challenging. Closed-form expressions are only known for Haar-random fermionic Gaussian states without~\cite{lydzba_rigol_20, bianchi_hackl_21} and with~\cite{bianchi_hackl_22} $U(1)$ symmetry, though analytic bounds have also been calculated for translationally invariant fermionic Gaussian states~\cite{vidmar_hackl_17, hackl_vidmar_19}.

%%%%%%%%%%%%%%%%%%%%%%
\begin{figure}[!t]
    \includegraphics[width=\columnwidth]{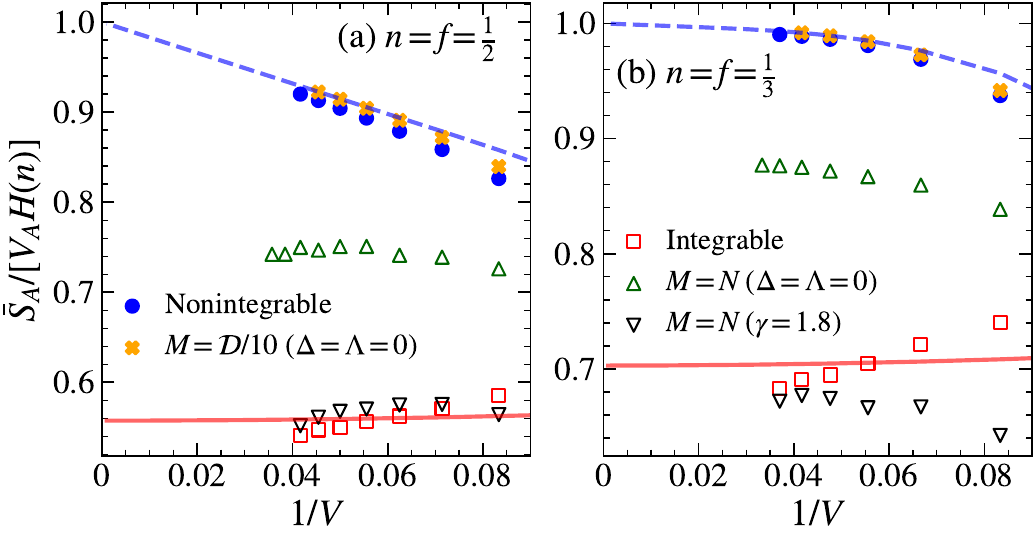}
    \vspace{-0.5cm}
    \caption{Normalized average entanglement entropy $\bar S_A$ vs $1/V$ in eigenstates of integrable (nonintegrable) interacting fermions with $\Lambda\!=\!0$ ($\Lambda\!=\!2$) for (a) $n\!=\!\frac{1}{2}$ and (b) $n\!=\!\frac{1}{3}$. The subsystem fraction is fixed at $f\!=\!V_A/V\!=\!n$. The solid and dashed lines are the predictions for Haar-random fermionic Gaussian states [Eq.~\eqref{eq:entanglement:gaussian:u1}] and Haar-random states [Eq.~\eqref{eq:page:u1}] with $U(1)$ symmetry~\cite{bianchi_hackl_22}, respectively. The empty triangles show results for $\ev{S^{}_A}^{}_{\bb=\frac12}$ for superpositions of $M\!=\!N$ (up and down triangles) and $M\!\approx\!0.1\mathcal{D}$ (crosses) Gaussian states constructed from eigenstates of $\hat H$~\eqref{eq:model} for $\Delta\!=\!\Lambda\!=\!0$ (up and crosses) and of $\hat{H}^{}_\text{RP}$~\eqref{sm:eq:rp} for $\gamma=1.8$ (down triangles).}\label{figM1}
\end{figure}
%%%%%%%%%%%%%%%%%%%%%%

Here we compute $\ev{S^{}_A}^{}_{\bb=\frac12}$ and $\bar S^{}_A$ numerically. In Fig.~\ref{figM1} we show $\ev{S^{}_A}^{}_{\bb=\frac12}/[V^{}_A H(n)]$ for $n\!=\!\tfrac12$ [Fig.~\ref{figM1}(a)] and $n\!=\!\tfrac13$ [Fig.~\ref{figM1}(b)] for $M\!=\!N$ (up and down triangles) and $M\!\approx\!0.1\mathcal{D}$ (crosses). The results for $M\!\approx\!0.1\mathcal{D}$ extrapolate to $1$ in the thermodynamic limit. In contrast, the results for $M\!=\!N$ (computed for two different bases of fermionic Gaussian states $\{\ket{\uppsi_{\alpha}^G}\}$ introduced later) extrapolate to a value smaller than 1 that depends on the basis used, as we discuss in more detail in what follows.

%%%%%%%%%%%%%%%%%%%%%%%%
\textit{Interacting spinless-fermions.}
We consider a particle-number-conserving model of interacting fermions in a 1D lattice with $V$ sites and open boundary conditions:
\begin{align}\label{eq:model}
    \hat{H}&=\hat{H}^{}_\text{n}+\Lambda\hat{H}^{}_\text{nn}+\epsilon^{}_1\hat{n}^{}_1+\epsilon^{}_L\hat{n}^{}_L,\quad \text{where}\nonumber\\
    \hat{H}^{}_\text{n}&=\sum_{\ell=1}^{V-1}\qty(\hat{c}_\ell^\dagger\hat{c}^{}_{\ell+1}+{\rm h.c.})+\Delta\sum_{\ell=1}^{V-1}\hat{n}^{}_\ell\hat{n}^{}_{\ell+1},\nonumber\\
    \hat{H}^{}_\text{nn}&=\sum_{\ell=1}^{V-2}\qty(\hat{c}_\ell^\dagger\hat{c}^{}_{\ell+2}+{\rm h.c.})+\hat{n}^{}_\ell\hat{n}^{}_{\ell+2},
\end{align}
For $\Delta\!=\!\Lambda\!=\!0$, $\hat{H}$ describes noninteracting fermions, and for $\Delta\!\neq\!0$ and $\Lambda\!=\!0$ the model is integrable via the Bethe Ansatz~\cite{Bethe_31, Baxter1972b}. On the other hand, when $\Lambda\!\neq\!0$ the Hamiltonian is nonintegrable. We choose $\Delta\!=\!0.55$ for all our calculations, and $\Lambda\!=\!2$ for the nonintegrable case (see Ref.~\cite{SM}). $\epsilon^{}_1$ and $\epsilon^{}_V$ are Sklyanin boundary fields that break the discrete symmetries of $\hat{H}^{}_\text{n}+\Lambda\hat{H}^{}_\text{nn}$ while preserving the integrability when $\Lambda\!=\!0$~\cite{Alcaraz1987, Sklyanin1988, MezincescuNepomechie1991, deVega1992, Jimbo1994, Batchelor1996, Nepomechie2003, Pozsgay2018}. We choose $\epsilon^{}_1\!=\!0.4$ and $\epsilon^{}_V\!=\!-0.6$. In all our numerical calculations, the averages are computed over $\min(500,\mathcal{D}/10)$ central eigenstates of the energy spectrum to reduce finite-size effects in the smallest chains.

In Fig.~\ref{figM2}, we plot the numerical results for $\bar{\mathcal{P}}-(2n-1)^2$ versus $1/V$ in the integrable (open squares) and nonintegrable (filled circles) regimes for $n\!=\!\frac{1}{2}$ [Fig.~\ref{figM2}(a)] and $n\!=\!\frac{1}{3}$ [Fig.~\ref{figM2}(b)]. In the main panels, the results in the nonintegrable regime are vanishingly small, and the insets show that this is because they decay polynomially with $\mathcal{D}$ for $\mathcal{D}\in[10^3,10^5]$, as predicted by Eq.~\eqref{eq:purity} for $M\!\propto\!D$ (see dashed lines in the insets). For $\mathcal{D}\!>\!10^5$ we find that the convergence becomes slower, likely because of slower corrections present in the Hamiltonian eigenstates. The main panels show that, on the other hand, the results in the integrable regime are well described by Eq.~\eqref{eq:purity} with $M\!\approx\!N$. 

The comparison between numerical results for $H(n)-\bar{\mathcal{N}}$ in Fig.~\ref{figM3} and the result for random superpositions of fermionic Gaussian states~\eqref{eq:non-gauss} is qualitatively similar to that in Fig.~\ref{figM2}. For both integrable and nonintegrable systems $\bar{\mathcal{P}}$ and $\bar{\mathcal{N}}$ approach the (maximally mixed-state) bounds found analytically, but with qualitatively different scalings of $M$ with system size. Together, the results in Figs.~\ref{figM2} and~\ref{figM3} support the conclusion that $\bar{\mathcal{P}}$ and $\bar{\mathcal{N}}$ in eigenstates of integrable (nonintegrable) models can be described using random superpositions of $M\!\propto\!N$ ($M\!\propto\!\mathcal{D}$) fermionic Gaussian states. Those comparisons, however, do not allow us to determine which type of fermionic Gaussian states form the superpositions.

In Figs.~\ref{figM1}(a) and~\ref{figM1}(b) we plot numerical results for $\bar{S}^{}_A/[V^{}_A H(n)]$ vs $1/V$ in the integrable (open squares) and nonintegrable (filled circles) regimes for $n\!=\!\tfrac12$ [Fig.~\ref{figM1}(a)] and $n\!=\!\tfrac13$ [Fig.~\ref{figM1}(b)]. The results in the nonintegrable regime increase with increasing $V$, and approach the value 1 predicted for Haar-random states at fixed particle filling $n$ in the thermodynamic limit. There are two sets of results that closely track those in the nonintegrable regime: (i) the exact analytical prediction of Eq.~\eqref{eq:page:u1} for Haar-random states in finite systems (dashed line), and (ii) the average $\ev{S^{}_A}^{}_{\bb=\frac12}$ over random superpositions of $M\!\approx\!0.1\mathcal{D}$ Gaussian states that are eigenstates of Hamiltonian~\eqref{eq:model} for $\Delta\!=\!\Lambda\!=\!0$ (crosses).

The results in the integrable regime, on the other hand, decrease as $V$ increases and approach a value smaller than 1 in the thermodynamic limit. This behavior is qualitatively similar to the analytical prediction of Eq.~\eqref{eq:entanglement:gaussian:u1} for Haar-random fermionic Gaussian states in finite systems (solid line) at fixed filling $n$. However, the results in the integrable regime extrapolate to a smaller value in the thermodynamic limit (see Fig.~\ref{figA2} for the results of extrapolations). Similar results have been obtained in earlier studies~\cite{leblond_mallayya_19, patil_rigol_23, swietek_kliczkowski_24}. In Fig.~\ref{figM1}, we also show numerical averages $\ev{S^{}_A}^{}_{\bb=\frac12}$ for $M\!=\!N$ taking as basis of Gaussian states the eigenstates of the Hamiltonian~\eqref{eq:model} in the noninteracting limit ($\Delta\!=\!\Lambda\!=\!0$). Those results appear to extrapolate to a value smaller than 1 in the thermodynamic limit. However, superpositions of $M\!=\!N$ of those Gaussian states lead to values of the average entanglement entropy that are higher than in the individual Gaussian states, which already have a higher entropy than the interacting eigenstates (see Fig.~\ref{figA2}).

Therefore, we conclude that the fermionic Gaussian states that are part of the $M=N$ superpositions in Eq.~\eqref{eq:superposition} that describe integrable states must have a lower entropy than in the noninteracting limit, while still remaining delocalized. Motivated by this finding, we also consider superpositions of eigenstates of the Rosenzweig-Porter (RP) model~\cite{rosenzweig_porter1960, kravtsov_khaymovich2015, Pino_2019, menzler_meisner_2026, venturelli_tarzia_23} defined as:
\begin{equation}\label{sm:eq:rp}
    \hat{H}^{}_\text{RP} = \sum_{\ell=1}^V\varepsilon^{}_\ell \hat{c}_\ell^\dagger\hat{c}^{}_\ell + V^{-\gamma/2}\sum_{\ell,\ell'=1}^V R^{}_{\ell\ell'}\hat{c}^\dagger_\ell\hat{c}^{}_{\ell'}\,,
\end{equation}
where $\varepsilon^{}_\ell$ are real Gaussian random variables with zero mean and unit variance, and ${R}^{}_{\ell\ell'}$ are elements of a GOE matrix, \ie ${R}^{}_{\ell\ell'}$ are sampled from a real Gaussian distribution with zero mean and variance $\overline{R^{2}_{\ell\ell}}\!=\!2$ and $\overline{R^{2}_{\ell\ell'}}\!=\!1$ ($\ell\neq\ell'$). The parameter $\gamma$ drives the model across two single-particle transitions: from an ergodic (small $\gamma$) to a fractal phase at $\gamma\!=\!1$~\cite{kravtsov_khaymovich2015, Pino_2019, menzler_meisner_2026} and from the fractal into a localized phase at $\gamma\!=\!2$~\cite{barney_galitski_23, cadez_dietz_24, kutlin2024investigating, swietek_vidmar2026}. We focus on the eigenstates for $1\!<\!\gamma\!<\!2$. In this regime, the fractal dimension of the eigenstates is known analytically with a value $d^{}_2\!=\!2-\gamma$~\cite{truong_2016, bogomolny_sieber_18}. We tuned the parameter $\gamma$ and found that, for $\gamma=1.8$, the results for $\ev{S^{}_A}^{}_{\bb=\frac12}$ are close to those obtained for the integrable Hamiltonian eigenstates (see down triangles in Fig.~\ref{figM1}).

%%%%%%%%%%%%%%
\textit{Conclusions.}
We derive closed-form expressions for the average one-body purity $\ev{\mathcal{P}}^{}_\bb$ and the non-Gaussianity $\ev{\mathcal{N}}^{}_\bb$ of random superpositions of $M$ fermionic Gaussian states with fixed particle number. We find that those two quantities, which are independent of the bases of Gaussian states used in the superpositions, can distinguish superpositions whose Gaussian rank scales as $M=O(N)$ and $M=O(\mathcal{D})$ as one increases the system size at fixed filling $n=N/V$. We studied numerically the corresponding averages $\bar{\mathcal{P}}$ and $\bar{\mathcal{N}}$ in eigenstates of integrable and nonintegrable regimes of $\hat H$~\eqref{eq:model} finding that they are consistent with the results for $M=O(N)$ and $M=O(\mathcal{D})$ superpositions of Gaussian states, respectively. We also find that random superpositions of fermionic Gaussian states built from noninteracting fermions in a fractal phase closely describe the average entanglement entropy of eigenstates in the integrable regime.

Since fermionic Gaussian states constitute a natural class of efficiently simulable fermionic states~\cite{valiant_quantumcomputersthat_2001, terhal_divincenzo_2002, weedbrook_gaussianquantuminformation_2012, bravyi_gosset_2017, majsak_simpleefficientjoint_2025}, our results suggest that typical highly excited eigenstates of interacting integrable models, despite being volume-law entangled as well as strongly interacting and therefore non-Gaussian, admit compact polynomial-rank representations with potential implications for efficient classical simulation~\cite{terhal_divincenzo_2002, bravyi_gosset_2017, liu_winter_2022} and shallow-circuit state preparation~\cite{motta_determiningeigenstatesthermal_2020, mcardle_quantumcomputationalchemistry_2020, joven_scalablequantumcomputational_2026}. Furthermore, our results for the entanglement entropy highlight the need to study ``fractal'' fermionic Gaussian states and their superpositions to gain an analytical understanding of entanglement in eigenstates of integrable interacting models that parallels that for eigenstates of nonintegrable models in terms of Haar-random states.

%%%%%%%%%%%%%%
\textit{Note added.--}
We have coordinated the submission of this manuscript with another independent related work~\cite{tarabunga2026}, which studies two families of fermionic non-Gaussianity measures, including a measure studied here, proves resource-theoretic properties, and apply them to ground states of systems without particle conservation.

%%%%%%%%%%%%%%
\textit{Acknowledgments.}
We acknowledge discussions with P.~Łydżba and W.~Dobrautz. R.S.~and L.V.~acknowledge support from the Slovenian Research and Innovation Agency (ARIS), Research core funding Grants No.~P1-0044, No.~N1-0273, No.~J1-50005, and No.~N1-0369, as well as the Consolidator Grant Boundary-101126364 of the European Research Council (ERC). M.R.~acknowledges support from the United States National Science Foundation (NSF) Grant No.~PHY-2309146, and M.K.~from the National Science Centre (Poland) Grant No.~2024/53/B/ST3/02756 (M.K.). We gratefully acknowledge the High Performance Computing Research Infrastructure Eastern Region (HCP RIVR) consortium~\cite{vega1} and European High Performance Computing Joint Undertaking (EuroHPC JU)~\cite{vega2} for funding this research by providing computing resources of the HPC system Vega at the Institute of Information sciences~\cite{vega3}.

\bibliography{references}

\clearpage
\newpage

\onecolumngrid
\begin{center}
{\large \bf End Matter}\\
\end{center}
\twocolumngrid

% \appendix

%%%%%%%%%%%%%%%%%%%%%%%%%%%%%%%%%%%%%%%%
{\it Appendix A: One-body purity.}
The average one-body purity $\ev{\mathcal{P}}^{}_\bb$ in Eq.~\eqref{eq:purity:def} of the complex structure ${\bf J}_{\Psi}$ in Eq.~\eqref{eq:obdm:mixed} can be computed exactly. We can write:
\begin{equation}
        \ev{\mathcal{P}}^{}_\bb = 1-4n+\frac{4}{V}\sum_{\alpha^{}_1,...,\alpha^{}_4}A^{\alpha^{}_1\alpha^{}_3}_{\alpha^{}_2\alpha^{}_4}B^{\alpha^{}_1\alpha^{}_3}_{\alpha^{}_2\alpha^{}_4}\,,
\end{equation}
where we defined the tensors
\begin{equation}
    A^{\alpha^{}_1\alpha^{}_3}_{\alpha^{}_2\alpha^{}_4} = \expval{a_{\alpha^{}_1}^*a^{}_{\alpha^{}_2}a_{\alpha^{}_3}^*a^{}_{\alpha^{}_4}}_\bb\;,
\end{equation}
and
\begin{equation}
    B^{\alpha^{}_1\alpha^{}_3}_{\alpha^{}_2\alpha^{}_4} =\sum_{q,q'}\overline{\mel{\uppsi^G_{\alpha^{}_1}}{\hat{f}_q^\dagger\hat{f}^{}_{q'}}{\uppsi^G_{\alpha^{}_2}}\mel{\uppsi^G_{\alpha^{}_3}}{\hat{f}_{q}^\dagger\hat{f}^{}_{q'}}{\uppsi^G_{\alpha^{}_4}}}\;.
\end{equation}
We can evaluate the sum over the four indices $\alpha^{}_1,...,\alpha^{}_4$ by dividing the sum into integer partitions of the form $4\!=\!3+1\!=\!2+2\!=\!2+1+1\!=\!1+1+1+1$ (each number represents an index that is distinct from all others). Due to the rotational invariance of the distribution of $a^{}_\alpha$, only terms that pair together $a^{}_\alpha$ and $a_\alpha^*$ contribute. Hence, the only nonvanishing terms are $A^{\alpha\alpha}_{\alpha\alpha}$, $A^{\alpha\beta}_{\alpha\beta}$, and $A^{\alpha\beta}_{\beta\alpha}$. This leads to the formula for purity
\begin{align}\label{eq:purity:exact_form}
\ev{\mathcal{P}}^{}_\bb\! =& 1 - 4n + 4n \!\sum_\alpha\! \expval{|a^{}_\alpha|^4}_\bb \!+\!\frac{4}{V} \!\sum_{\substack{\alpha\neq\beta\\ q}}\! \expval{|a^{}_\alpha|^2|a^{}_\beta|^2}_\bb\! \overline{n_q^\alpha n_q^\beta}\nonumber\\
+&\frac{4}{V}\sum_{\substack{\alpha\neq\beta\\ q,q'}}\expval{|a^{}_\alpha|^2|a^{}_\beta|^2}_\bb \overline{\left|{\mel{\uppsi^G_{\alpha}}{\hat{f}_q^\dagger\hat{f}^{}_{q'}}{\uppsi^G_{\beta}}}\right|^2}.
\end{align}
Since the second sum involves distinct indices $\alpha\!\neq\!\beta$, we can write $\overline{n_q^\alpha n_q^\beta} = \overline{n_q^\alpha}\overline{n_q^\beta}=n^2$. The last term can be evaluated by noticing that ${\mel{\uppsi^G_{\alpha}}{\hat{f}_q^\dagger\hat{f}^{}_{q'}}{\uppsi^G_{\beta}}}^2\!=\!X_{\alpha\beta}$ can be treated as a random variable whose value is $1$ with probability $p$ that the two states differ by two orbitals. For any given state $\ket{\uppsi^G_\alpha}$ there are $N(V-N)$ states $\ket{\uppsi^G_\beta}\!\neq\!\ket{\uppsi^G_\alpha}$ that differ by two orbitals (out of $\mathcal{D}$ states in total). Therefore, the probability $p$ is:
\begin{equation}\label{eq:prob:index_pair}
    p=\frac{N(V-N)}{\mathcal{D}}\,.
\end{equation}
This allows us to compute the expectation value of $X_{\alpha\beta}$ for superpositions of $M$ fermionic Gaussian states. The number of states that differ by two orbitals is $X\!=\!\sum_{\alpha\neq\beta}X_{\alpha\beta}$, with an expectation value equal to
\begin{equation}\label{eq:prob:average_num_states}
    \mathbb{E}[X] = \sum_{\alpha\neq\beta}\mathbb{E}[X_{\alpha\beta}] = M(M-1)\frac{N(V-N)}{\mathcal{D}}\;.
\end{equation}
The average over the random coefficients in Eq.~\eqref{eq:purity:exact_form} is discussed in Ref.~\cite{SM}, see~Eqs.~\eqref{sm:eq:dirichlet:moments} and~\eqref{sm:eq:dirichlet:monomials}. Combining these results with Eq.~\eqref{eq:prob:average_num_states} yields the expression for the one-body purity in Eq.~\eqref{eq:purity}.

\begin{figure}[!t]
    \includegraphics[width=\linewidth]{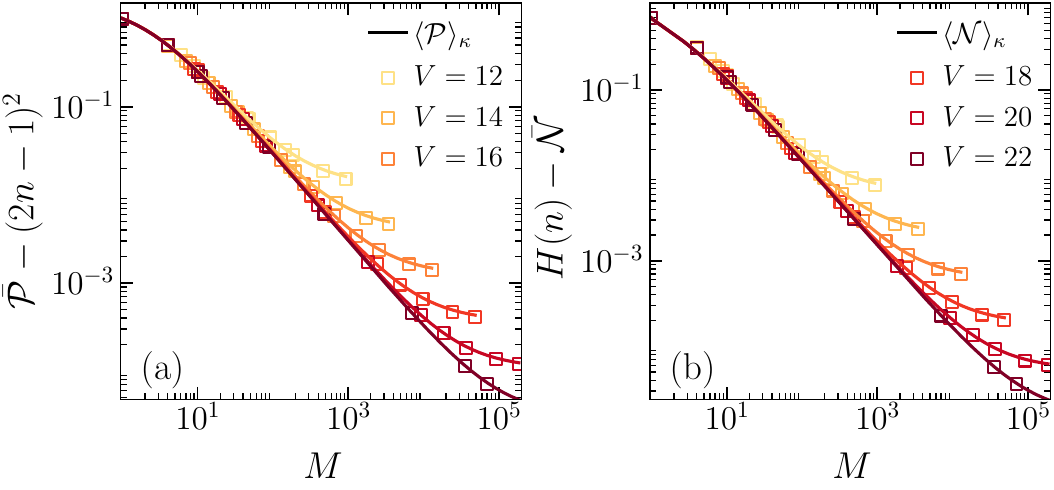}
    \vspace{-0.6cm}
    \caption{(a) Average one-body purity for random superpositions of $M$ eigenstates of the Hamiltonian~\eqref{eq:model} with $\Delta\!=\!\Lambda\!=\!0$ at $n\!=\!\tfrac12$. The solid lines are the analytic predictions of Eq.~\eqref{eq:purity}. (b) Average non-Gaussianity for the same superpositions as in (a). The solid lines are the analytic predictions of Eq.~\eqref{eq:non-gauss}.}\label{fig5}
\end{figure}

In Fig.~\ref{fig5}(a) we report numerical results (open squares) for the average purity of random superpositions of eigenstates of the Hamiltonian~\eqref{eq:model} in the noninteracting limit ($\Delta\!=\!\Lambda\!=\!0$) at $n\!=\!1/2$. They are described by Eq.~\eqref{eq:purity}, whose predictions are shown as solid lines.

%%%%%%%%%%%%%%%%%%%%%%%%%%%%%%%%%%%%%%%%
{\it Appendix B: Non-Gaussianity.}
To compute the non-Gaussianity, we assume that:
\begin{equation}\label{eq:higher_order:approx}
    \frac{1}{V}\overline{\Tr({\bf J}_\Psi)^{2k}} \approx \qty[\frac{1}{V}\overline{\Tr({\bf J}_\Psi)^2}]^k = \ev{\mathcal{P}}_\bb^k\;,
\end{equation}
where $\ev{\mathcal{P}}^{}_\bb$ is given by Eq.~\eqref{eq:purity}. The validity of this assumption is tested in Ref.~\cite{SM}.

Under this assumption, the non-Gaussianity can be written as:
\begin{equation}\label{eq:non-gaussianity:approx}
        \ev{\mathcal{N}}^{}_\bb = \ln{2} - \sum_{k=1}^\infty\frac{\ev{\mathcal{P}}_\bb^k}{2k(2k-1)},
\end{equation}
whose closed-form expression is given in Eq.~\eqref{eq:non-gauss}.

In Fig.~\ref{fig5}(b) we show numerical results (open squares) for the average non-Gaussianity of random superpositions of eigenstates of $\hat H$~\eqref{eq:model} in the noninteracting limit ($\Delta\!=\!\Lambda\!=\!0$) at $n\!=\!1/2$. They are described by Eq.~\eqref{eq:non-gauss}, whose predictions are shown as solid lines.

%%%%%%%%%%%%%%%%%%%%%%%%%%%%%%%%%%%%%%%%%%%%%%%%%%%
{\it Appendix C: Entanglement entropy.}
The average entanglement entropy of Haar-random pure states has the form~\cite{Page93, bianchi_hackl_22}:
\begin{equation}\label{eq:page}
    \expval{S_A}=\Psi(\mathcal{D}_A\mathcal{D}_B+1)-\Psi(\mathcal{D}_B+1)-\frac{\mathcal{D}_A-1}{2\mathcal{D}_B}\;,
\end{equation}
for $\mathcal{D}_A\!<\!\mathcal{D}_B$, where $\Psi(x)\!=\!\Gamma'(x)/\Gamma(x)$ is  the digamma function. The expression for $\mathcal{D}_A\!>\!\mathcal{D}_B$ is obtained by exchanging $\mathcal{D}_A\leftrightarrow\mathcal{D}_B$ in Eq.~\eqref{eq:page}. For qubit-based Hilbert spaces, $\mathcal{D}_A\!=\!2^{V^{}_A}$, and for $V^{}_A\!\gg\!1$ and $V^{}_B\!\gg\!1$, the leading nonvanishing terms read:
\begin{equation}
    \expval{S_A} = V^{}_A\ln{2} - \frac{1}{2}\delta^{}_{f,\frac{1}{2}}\,,\ \ \text{for}\ \ f=\frac{V^{}_A}{V}\leq\frac12.
\end{equation}

In the presence of $U(1)$ symmetry, for a fixed particle number $N$, the Hilbert space can be decomposed as:
\begin{equation}
    \mathcal{H}^{(N)} = \bigoplus_{N_A=0}^{{\rm min}(N, V^{}_A)}\qty(\mathcal{H}_A^{(N_A)}\otimes\mathcal{H}_B^{(N-N_A)})\;,
\end{equation}
where we assume $N\!\leq\!V/2$. Again, for qubit-based systems, the dimensions of the subsystem Hilbert spaces are:
\begin{equation}
    \mathcal{D}_i^{(N_i)}={\rm dim}\mathcal{H}_i^{(N_i)}=\binom{V^{}_i}{N_i}
\end{equation}
for $i\!=\!A,B$, and the full Hilbert space is $\mathcal{D}\!=\!\binom{V}{N}$. The average entanglement entropy of Haar-random pure states with fixed particle number reads~\cite{bianchi_dona_19, bianchi_hackl_22}
\begin{align}\label{eq:page:u1}
        \expval{S_A}^{}_N = \sum_{N_A=0}^{{\rm min}(N, V^{}_A)}&\frac{\mathcal{D}_A^{(N_A)}\mathcal{D}_B^{(N_B)}}{\mathcal{D}}\Big[\expval{S_A}+\Psi(\mathcal{D}+1)\nonumber \\ &-\Psi(\mathcal{D}_A^{(N_A)}\mathcal{D}_B^{(N_B)}+1)\Big],
\end{align}
where $\expval{S_A}$ is the Page prediction in Eq.~\eqref{eq:page} for subsystem dimensions $\mathcal{D}_A^{(N_A)}$ and $\mathcal{D}_B^{(N_B)}$. In the large subsystems limit, the leading volume-law term reads
\begin{equation}
    \expval{S_A}_N \simeq -[n\ln(n)+(1-n)\ln(1-n)]\,V^{}_A = H(n)\,V^{}_A\,,
\end{equation}
where $H(n)$ is the bound for $\ev{\mathcal{N}}^{}_\bb$ in Eq.~\eqref{eq:non-gauss:leading}.

On the other hand, for Haar-random fermionic Gaussian states the average entanglement entropy reads~\cite{bianchi_hackl_21, bianchi_hackl_22}
\begin{align}\label{eq:entanglement:gaussian}
    \expval{S_A}^{}_G =& \qty(V-\frac{1}{2})\Psi(2V) + \qty(\frac{1}{2} - V^{}_B)\Psi(2V^{}_B)\nonumber\\ &+\qty(\frac{1}{4}-V^{}_A)\Psi(V)-\frac{1}{4}\Psi(V^{}_B)-V^{}_A\,.
\end{align}
In the presence of $U(1)$ symmetry, for a fixed particle number $N$, the average reads~\cite{bianchi_hackl_22}:
\begin{align}\label{eq:entanglement:gaussian:u1}
    \expval{S_A}^{}_{G,N} =& 1 - \frac{V^{}_A}{V}\qty[(V\!-\!N)\Psi(V\!-\!N) + N\Psi(N) + V+1 ]\nonumber\\ &+V\Psi(V)-V^{}_B\Psi(V^{}_B\!+\!1)\;,
\end{align}
for $V^{}_A\!\leq\! N\!\leq\! V/2$, which we consider here. The expressions for all the remaining particle numbers and subsystem sizes follow straightforwardly, see Ref.~\cite{bianchi_hackl_22}.

\begin{figure}[!t]
    \includegraphics[width=\columnwidth]{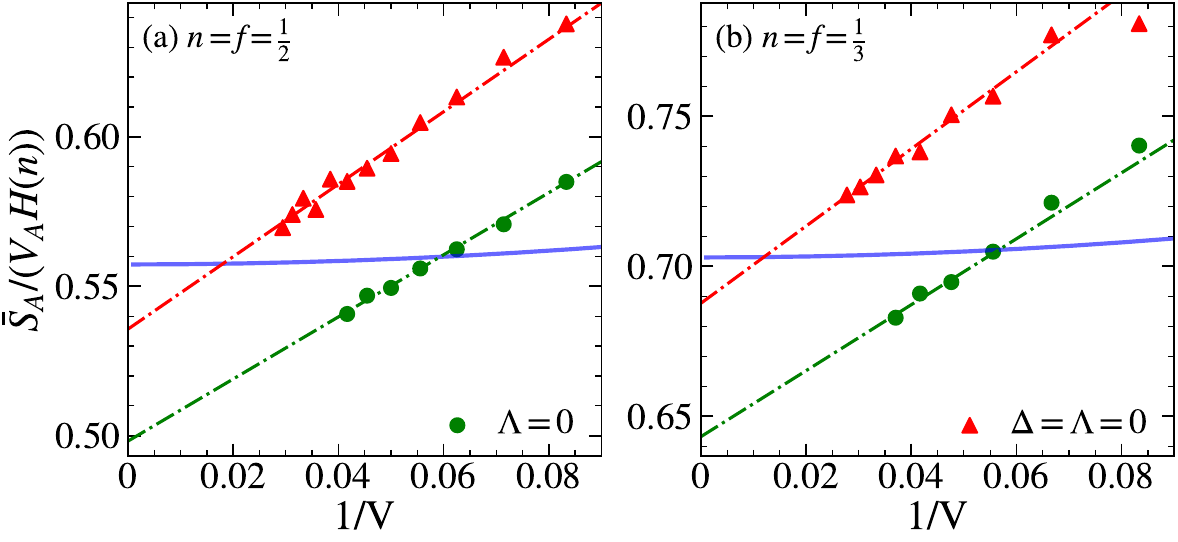}
    \vspace{-0.6cm}
    \caption{Average entanglement entropy for integrable-interacting ($\Lambda\!=\!0$, shown in the main text) and noninteracting ($\Delta\!=\!\Lambda\!=\!0$) fermions for (a) $n\!=\!f\!=\!\frac{1}{2}$, (b) $n\!=\!f\!=\!1/3$. The dash-dotted lines show fits of the numerical results (excluding the three smallest system sizes) to $s^{}_{\infty}+c/V$. The solid lines show the predictions of Eq.~\eqref{eq:entanglement:gaussian:u1}.}\label{figA2}
\end{figure}

There are no closed-form expressions for the average entanglement entropy of other types of fermionic Gaussian states, though numerical results and tight analytic bounds that are qualitatively similar to the predictions of Eqs.~\eqref{eq:entanglement:gaussian} and~\eqref{eq:entanglement:gaussian:u1} have been obtained for translationally invariant fermionic Gaussian states~\cite{vidmar_hackl_17, hackl_vidmar_19}. Among the quantitative differences identified is the fact that at $n\!=\!f\!=\!1/2$, $\expval{S_A}^{}_{G,N}/(V^{}_A\ln 2) \!=\! \expval{S_A}^{}_{G}/(V^{}_A\ln 2) \!=\! 2\!-\!(\ln{2})^{-1} \!\approx\! 0.5573$, while the numerically extrapolated result for translationally invariant free fermions is smaller $\bar{S}_A/(V^{}_A\ln 2) \!=\! 0.5378(1)$~\cite{vidmar_hackl_17, bianchi_hackl_22}.

In Fig.~\ref{figA2}, we compare the predictions of Eq.~\eqref{eq:entanglement:gaussian:u1} and the numerical results obtained for the average over eigenstates of $\hat H$~\eqref{eq:model} in the interacting integrable ($\Lambda\!=\!0$) and noninteracting ($\Delta\!=\!\Lambda\!=\!0$) regimes. The results in the noninteracting limit extrapolate to values that are below the predictions of Eq.~\eqref{eq:entanglement:gaussian:u1}, as found for translationally invariant fermionic Gaussian states. The results for the interacting integrable model extrapolate to even smaller values. Our study suggests that the latter is the case because the typical interacting eigenstates can be modeled by superpositions of fractal fermionic Gaussian states.

%% so references appear in the list that are in the supplemental when it is split for PRL submission
\nocite{mehta_91,mehta2004,Joe1997,devroye1986,Oganesyan07,Atas_13}
\clearpage
%\end{document}

\newpage
\phantom{a}
\newpage
%%%%%%%%%%%%%%%%%%%%%%%%%%%%%%%%%%%%%%%%

\onecolumngrid

\begin{center}

{\large \bf Supplemental Material: One-Body~Purity,~Non-Gaussianity,~and~Entanglement~in~Interacting~Integrable~Models}\\

\begin{center}
R. Świętek$^{1,2,3,4}$, M. Kliczkowski$^{5,6}$, L. Vidmar$^{2,3}$, and M. Rigol$^{4}$\\
$^1${\it Institut f\"ur Theoretische Physik, Georg-August-Universität G\"ottingen, D-37077 G\"ottingen, Germany}\\
$^2${\it Department of Theoretical Physics, J. Stefan Institute, SI-1000 Ljubljana, Slovenia} \\
$^3${\it Department of Physics, Faculty of Mathematics and Physics, University of Ljubljana, SI-1000 Ljubljana, Slovenia} \\
$^4${\it Department of Physics, The Pennsylvania State University, University Park, Pennsylvania 16802, USA}\\
$^5${\it Center for Advanced Systems Understanding, Helmholtz-Zentrum Dresden-Rossendorf, Germany}\\
$^6${\it Institute of Theoretical Physics, Faculty of Fundamental Problems of Technology, Wrocław University of Science and Technology, 50-370 Wrocław, Poland}

\end{center}

\vspace{0.3cm}

\setcounter{page}{1}

\end{center}

\makeatletter
\@removefromreset{equation}{section}
\makeatother

\vspace{0.6cm}

%\begin{bibunit}
\twocolumngrid

\label{pagesupp}
\setcounter{secnumdepth}{3}
\renewcommand{\thetable}{S\arabic{table}}
\renewcommand{\thefigure}{S\arabic{figure}}
\renewcommand{\theequation}{S\arabic{equation}}
\renewcommand{\thepage}{S\arabic{page}}

\renewcommand{\thesection}{S\arabic{section}}

\setcounter{figure}{0}
\setcounter{equation}{0}
\setcounter{table}{0}
\setcounter{section}{0}

%%%%%%%%%%%%%%%%%%%%%%%%%%%%%%%%%%%%%%%%
\section{Linear complex structure}
For a generic state $\ket{\Phi}$, all one-body correlations are encoded in the linear complex structure~\cite{hackl_bianchi_21}:
\begin{equation}\label{sm:eq:complex_structure}
    i\mathcal{J}_{\Phi}=
    \begin{pmatrix}
        \mathbf{J}^{}_{\Phi}& -\mathbf{G}_{\Phi}^*\\ \mathbf{G}^{}_{\Phi}& -\mathbf{J}_{\Phi}^*
    \end{pmatrix}\;,
\end{equation}
where $(J^{}_{\Phi})^{}_{\ell\ell'}=\mel{\Phi}{[\hat{c}_\ell^\dagger,\hat{c}^{}_{\ell'}]}{\Phi}$ and $(G_{\Phi})^{}_{\ell\ell'}=\mel{\Phi}{[\hat{c}^{}_\ell,\hat{c}^{}_{\ell'}]}{\Phi}$. If the state is a fermionic Gaussian state $\ket{\Phi}=\ket{\uppsi_{\alpha}^G}$, then the linear structure fully characterizes it. Here, however, we are also interested in the properties of the linear structure for states that are superpositions of fermionic Gaussian states $\ket{\Phi}=\ket{\Psi}$ [cf.~Eq.~\eqref{eq:superposition} in the main text] and in eigenstates of interacting Hamiltonians $\ket{\Phi}=\ket{\psi^{}_m}$.

Alternatively, one can encode all one-body correlations in the covariance matrix:
\begin{equation}
    (\Gamma_\Phi)^{}_{ab} = -\frac{i}{2}\mel{\Phi}{[\hat{\gamma}_a,\hat{\gamma}_{b}]}{\Phi}\;,
\end{equation}
where $a,b=1,...,2V$, and the Majorana operators are defined as
\begin{equation}
\gamma^{}_{2\ell-1}=\hat{c}^{}_\ell + \hat{c}^\dagger_\ell, \qquad \gamma^{}_{2\ell} = -i(\hat{c}^{}_\ell - \hat{c}^\dagger_\ell).
\end{equation}
The linear complex structure and the covariance matrix are related by a similarity transformation $\mathbf{\Gamma}_\Phi = \frac{1}{2}W\mathcal{J}^{}_\Phi W^\dagger$, where $WW^\dagger=W^\dagger W=2\mathbb{I}$.

In our work we consider only $U(1)$ symmetric models, for which the linear complex structure simplifies to
\begin{equation}\label{sm:eq:complex_structure:U1}
    i\mathcal{J}^{U(1)}_\Phi =
    \begin{pmatrix}
        \mathbf{J}_\Phi & 0\\ 0 & -\mathbf{J}_\Phi^*
    \end{pmatrix},
\end{equation}
which allows us to simplify higher-order traces of the linear complex structure to $\Tr([i\mathcal{J}^{U(1)}_\Phi]^{2k}) = 2\Tr[(\mathbf{J}_\Phi)^{2k}]$.

%%%%%%%%%%%%%%%%
\subsection{Fermionic antiflatness and one-body purity}

A recent work~\cite{sierant_turkeshi_26} studied the fermionic antiflatness:
\begin{equation}\label{sm:eq:faf}
    \mathcal{F}_k(\ket{\Phi}) = V - \frac{1}{2}\Tr[({\bf \Gamma}_\Phi^T{\bf \Gamma}^{}_\Phi)^k],
\end{equation}
which is defined in terms of the covariance matrix ${\bf \Gamma}_\Phi$ for a state $\Phi$. For Haar-random states it was shown that
\begin{equation}
    \mathcal{F}_{k=1}(\ket{\Phi})=V - \frac{V(2V-1)}{\mathcal{D}+1}\;.
\end{equation}

Using the similarity transformation mentioned before, we can express the fermionic antiflatness for $k=1$ as:
\begin{align}\label{sm:eq:antiflatness:with:J}
        \mathcal{F}_1(\ket{\Phi}) 
        &= V - \frac{1}{2}\Tr({\bf \Gamma}_\Phi^T{\bf \Gamma}^{}_\Phi)
        = V + \frac{1}{2}\Tr({\bf \Gamma}_\Phi^2)\\
        &= V + \frac{1}{8}\Tr[\qty(W\mathcal{J}^{}_\Phi W^\dagger)^2]
        = V + \frac{1}{2}\Tr[(\mathcal{J}^{}_\Phi)^2]\;,\nonumber
\end{align}
where we used the antisymmetry of the covariance matrix ${\bf \Gamma}_\Phi^T=-{\bf \Gamma}^{}_\Phi$, the cyclic property of the trace, and the the fact that $WW^\dagger=W^\dagger W=2\mathbb{I}$.

Hence, we can write
\begin{equation}\label{sm:eq:faf_to_purity}
    \mathcal{F}_1(\ket{\Phi}) = V - \Tr(\mathbf{J}_\Phi^2) - \Tr(\mathbf{G}_\Phi\mathbf{G}_\Phi^*)\;,
\end{equation}
where we used that $\Tr[(\mathbf{J}_\Phi^*)^2]=\Tr(\mathbf{J}_\Phi^2)$. Therefore, the fermionic antiflatness for $k=1$ can be directly linked to the purity of the matrix $\mathbf{J}_\Phi$ and the trace of the product of the pair-correlation matrix $\mathbf{G}_\Phi$ and its conjugate $\mathbf{G}_\Phi^*$.

% %%%%%%%%%%%%%%%%
For systems with $U(1)$ symmetry, for which $\mathbf{G}^{}_\Phi=\mathbf{G}_\Phi^*=\mathbf{0}$, the average fermionic antiflatness over random superpositions of $M$ fermionic Gaussian states is related to the one-body purity defined in the main text:
\begin{equation}
    \ev{\mathcal{F}_1(\ket{\Phi})}^{}_\bb = V(1-\ev{\mathcal{P}}^{}_\bb)\;.
\end{equation}
Therefore, our result for $\ev{\mathcal{P}}^{}_\bb$ determines $\ev{\mathcal{F}_1(\ket{\Phi})}^{}_\bb$ for such superpositions.

\section{Distribution of coefficients}

In the main text, we consider a set of i.i.d.~Gaussian random coefficients $\tilde{a}_\alpha$ that are real (GOE case) or complex (GUE case). The normalized vector 
\begin{equation}
\mathbf{a} = \mathbf{\tilde{a}} / \|\mathbf{\tilde{a}}\|
\end{equation}
is Haar-distributed on the unit sphere in $\mathbb R^M$ or $\mathbb C^M$, respectively. This follows from the spherical symmetry of the Gaussian distribution and the invariance of the Haar measure under orthogonal or unitary transformations~\cite{mehta_91, mehta2004}. For $M<D$, the Haar distribution refers to the unit sphere in the chosen $M$-dimensional subspace rather than to the full $D$-dimensional Hilbert space.

Let $\mathbf z\in\mathbb F^M$ have i.i.d.~standard Gaussian entries, where $\mathbb F=\mathbb R$ or $\mathbb C$. Then the normalized vector $\mathbf z/\|\mathbf z\|$ is uniformly (equivalently, Haar) distributed on the unit sphere in $\mathbb F^M$. The squared magnitudes
\begin{equation}
x^{}_\alpha=|a_\alpha|^2=|z_\alpha|^2/\sum_{\beta=1}^M |z_\beta|^2
\end{equation}
are distributed as
\begin{equation}
    (x^{}_1,...,x^{}_M)\sim\mathrm{Dirichlet}(\bb, \dots, \bb)\;,
\end{equation}
with $\bb = 1/2$ for real Gaussians and $\bb = 1$ for complex Gaussians.

The $n$-th moment of $x^{}_\alpha$ in the Dirichlet distribution $\mathrm{Dirichlet}(\bb, \dots, \bb)$ with $M$ equal parameters is given by the standard formula \cite{Joe1997, devroye1986}:
\begin{equation}\label{sm:eq:dirichlet:moments}
    \mathbb{E}[(x^{}_\alpha)^n] = \frac{\prod_{j=0}^{n-1} (\bb + j)}{\prod_{j=0}^{n-1} (M\bb + j)} = \frac{(\bb)_n}{(M\bb)_n}
\end{equation}
where $(x)_n = x(x+1)\cdots(x+n-1)$ is the rising factorial (Pochhammer symbol). This expression is valid for all integers $n \!\geq\! 1$. More generally, the expectation value of a monomial in the components is given by
\begin{equation}\label{sm:eq:dirichlet:monomials}
    \mathbb{E}\qty[\prod_{\alpha=1}^M(x^{}_\alpha)^{p^{}_\alpha}] = \frac{\prod_{\alpha=1}^M\qty(\bb)_{p^{}_\alpha}}{\qty(M\bb)_P}\;,
\end{equation}
where we define $P=\sum_{\alpha=1}^M p^{}_\alpha$.

For completeness, the first few moments for the coefficients $|a^{}_\alpha|^2$ are
\begin{align}
    \mathbb{E}[|a^{}_\alpha|^2] &= \frac{\bb}{M\bb} = \frac{1}{M}, \\
    \mathbb{E}[|a^{}_\alpha|^4] &= \frac{\bb(\bb + 1)}{M\bb(M\bb + 1)} = \frac{\bb + 1}{M(M\bb + 1)}, \\
    \mathbb{E}[|a^{}_\alpha|^6] &= \frac{(\bb + 1)(\bb + 2)}{M(M\bb + 1)(M\bb + 2)},
\end{align}
while for $\alpha\neq\beta$ the first few moments of monomials are of the form
\begin{align}
\mathbb{E}\left[ |a^{}_\alpha|^2 |a^{}_\beta|^2 \right] &= \frac{ \bb^2 }{ M\bb (M\bb + 1) }\;, \\
\mathbb{E}\left[ |a^{}_\alpha|^2 |a^{}_\beta|^4 \right] &= \frac{ \bb^2 (\bb + 1) }{ M\bb (M\bb + 1)(M\bb + 2) }.
\end{align}
These results serve as the building blocks for deriving analytical expressions for the purity in Eq.~\eqref{eq:purity:def} and the non-Gaussianity in Eq.~\eqref{eq:non_gauss:def}.

\section{Maximally ergodic regime}

The Hamiltonian in Eq.~\eqref{eq:model} has a rich phase diagram with multiple integrable limits. After fixing $\Delta=0.55$, we looked for a value of the integrability breaking parameter $\Lambda$ around which the results are nearly independent of $\Lambda$. 

We computed (over the central 50\% of the energy spectrum) the ratio:
\begin{equation}\label{eq:gap_ratio}
    r^{}_\alpha=\frac{\min\qty{s^{}_\alpha,s^{}_{\alpha+1}}}{\max\qty{s^{}_\alpha,s^{}_{\alpha+1}}}\;,
\end{equation}
of consecutive level spacings $s^{}_\alpha\!=\!E_{\alpha+1}-E_\alpha$. The RMT prediction is $r^{}_{\rm GOE}\!=\!0.5307$~\cite{Oganesyan07,Atas_13}. Moreover, we computed the average eigenstate entanglement entropy $\bar S_A$, the non-Gaussianity [Eq.~\eqref{eq:non_gauss:def}] and the purity [Eq.~\eqref{eq:purity:def}] of the complex structure $\mathbf{J}_{\psi^{}_m}$ over $\min(500,\mathcal{D}/10)$ eigenstates in the center of the energy spectrum.

\begin{figure}[!h]
    \includegraphics[width=\columnwidth]{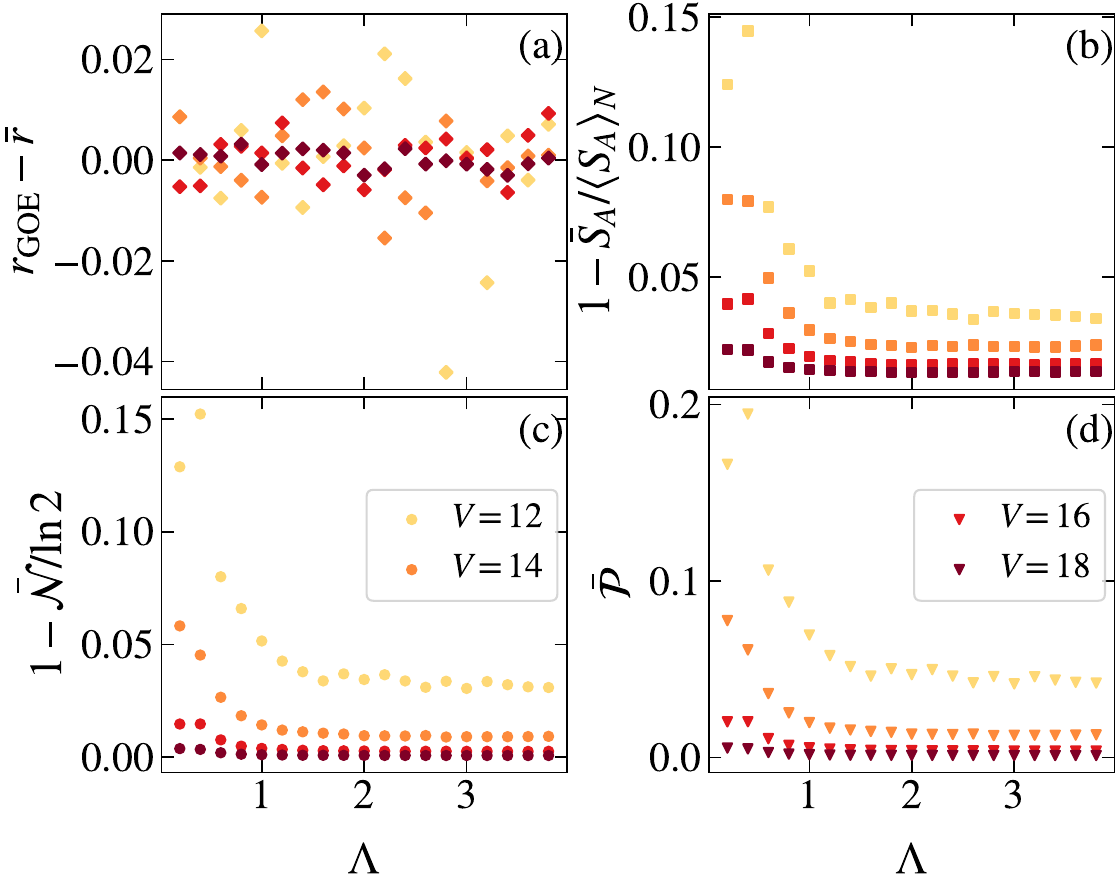}
    \vspace{-0.6cm}
    \caption{Various indicators of ergodicity: (a) ratio of consecutive level spacings, (b) entanglement entropy, (c) non-Gaussianity, (d) purity as functions of the integrability breaking parameter $\Lambda$.}\label{figA1}
\end{figure}

In Fig.~\ref{figA1} we show our numerical results. For $\Lambda\!>\!1$ one can see that a plateau develops, for all observables, for all but the smallest system size so we selected $\Lambda\!=\!2$ for our simulations.

%%%%%%%%%%%%%%%%%%%%%%%%%%%%%%%%%%%%%%%%

\section{Validity of Eq.~(\ref{eq:higher_order:approx})}

\begin{figure}[!b]
    \centering
    \includegraphics[width=\linewidth]{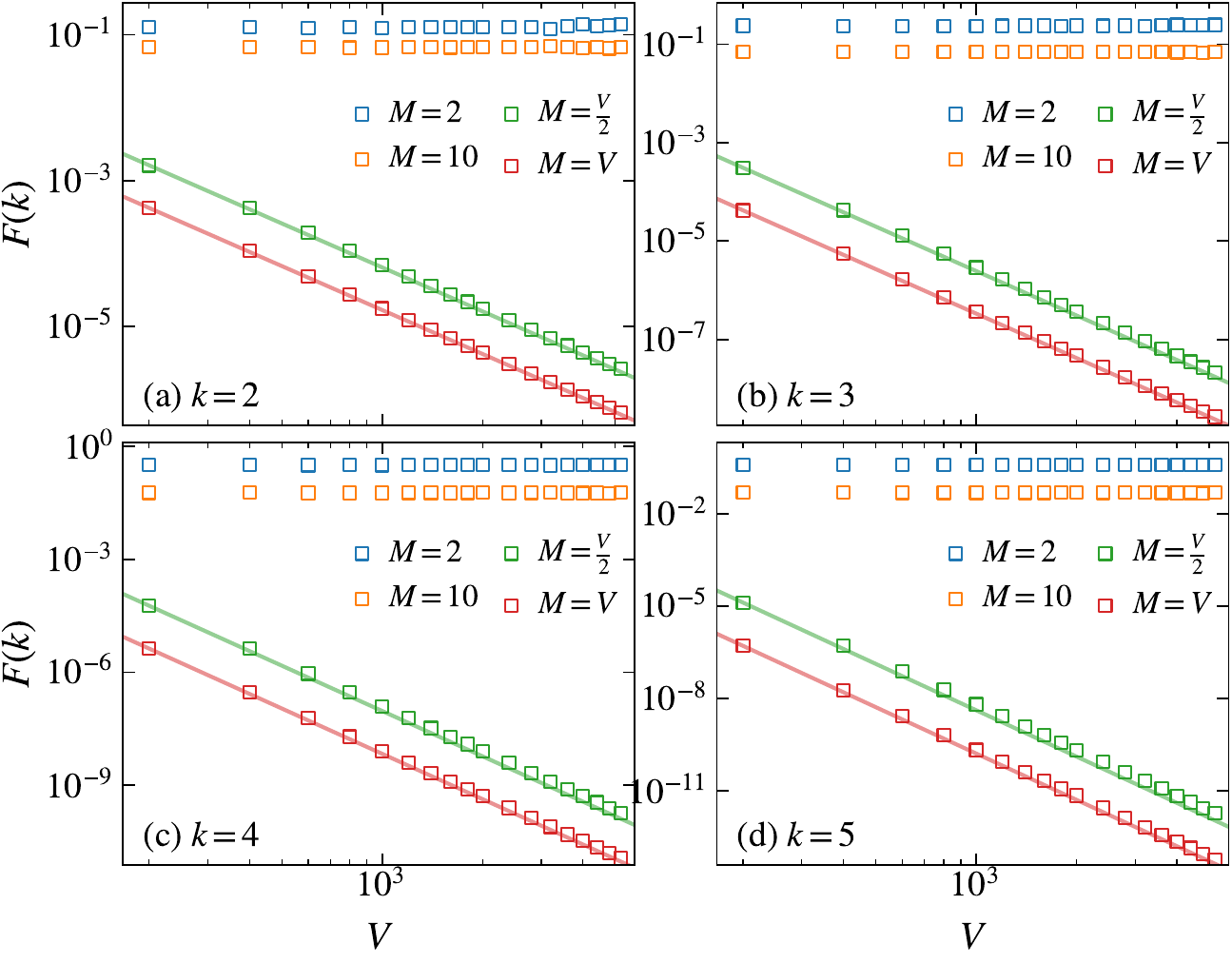}
    \vspace{-0.6cm}
    \caption{Absolute difference $F(k)$ in Eq.~\eqref{sm:eq:error} between the $k$-order trace and the expression for purity in Eq.~\eqref{eq:purity} to the $k$-th power for (a) $k=2$, (b) $k=3$, (c) $k=4$ and (d) $k=5$.
    The solid lines show a fit of $A/M^k$ to the numerical data.}\label{figS3}
\end{figure}

Here we test the validity of the approximation in Eq.~\eqref{eq:higher_order:approx}: 
\begin{equation}\label{sm:eq:error}
    F(k) = \abs{\frac{1}{V}\overline{\Tr[(\mathbf{J}_\Psi)^{2k}]} - \ev{\mathcal{P}}_\bb^k}\,.
\end{equation}
We consider many-body eigenstates of the SYK2 ensemble given by the Hamiltonian
\begin{equation}\label{sm:eq:syk2}
    \hat{H}^{}_{\rm SYK2} = \sum_{\ell,\ell'}A^{}_{\ell\ell'}\hat{c}_\ell^\dagger\hat{c}^{}_{\ell'}\,,
\end{equation}
where $A_{i,j}$ are elements of a GOE matrix. We show the results for $F(k)$ for $k\!\leq\!5$ for random superpositions of $M$ eigenstates as function of system size $V$ in Fig.~\ref{figS3}. We find that for fixed values of $M\!=\!O(1)$ the difference between the exact result and the approximation does not change with increasing $V$, and decreases only with increasing $M$. On the other hand, for $M\!=\!O(V)$ (which is the main interest in this work), the error of the approximation decays polynomially with $M$. We fit the numerical data to $A/M^k$ and find qualitatively good agreement. 

Therefore, we conclude that the approximation in Eq.~\eqref{eq:higher_order:approx} of the main text becomes exact in the thermodynamic limit for $M\!=\!O(V)$ and $M\!=\!O(\mathcal{D})$.

%%%%%%%%%%%%%%%%%%%%%%%%%%%%%%%%%%%%%%

\end{document}